\documentclass[conference]{IEEEtran} %!PN
\usepackage[dvips]{graphicx}
\usepackage{subcaption}
\usepackage{epsf}
\usepackage{epsfig}
\usepackage[noadjust]{cite}
\usepackage{amsmath}
\usepackage{url}
\usepackage[export]{adjustbox}
\usepackage{soul}
\usepackage{algorithm}
\usepackage[noend]{algpseudocode}
\usepackage{color}
\usepackage{tree-dvips}
\usepackage{booktabs}
\usepackage{multirow}
\usepackage{footnote}
\usepackage[T1]{fontenc}
\usepackage{mathtools, nccmath}
\usepackage[T1]{fontenc}

\DeclarePairedDelimiter\abs{\lvert}{\rvert}

\def\BibTeX{{\rm B\kern-.05em{\sc i\kern-.025em b}\kern-.08em
    T\kern-.1667em\lower.7ex\hbox{E}\kern-.125emX}}

\DeclarePairedDelimiter{\nint}\lfloor\rceil
\begin{document}

\onecolumn 

{\LARGE IEEE Copyright Notice} 
\\

\copyright 2020 IEEE. Personal use of this material is permitted. Permission from IEEE must be obtained for all other uses, in any current or future media, including reprinting/republishing this material for advertising or promotional purposes, creating
new collective works, for resale or redistribution to servers or lists, or reuse of any copyrighted component of this work in other works. \\

{\large Accepted to be Published in: Proceedings of the 21$^{st}$ International Symposium on Quality Electronic Design (ISQED 2020), Mar. 25-26, 2020, Santa Clara, CA.}

\twocolumn

\title{Analytical Estimation and Localization of Hardware Trojan Vulnerability in RTL Designs} %!PN

\author{\IEEEauthorblockN{Sheikh Ariful Islam, Love Kumar Sah, and Srinivas Katkoori}
\IEEEauthorblockA{Department of Computer Science and Engineering\\
University of South Florida \\
Tampa, FL 33620\\
Email: \{sheikhariful, lsah, katkoori\}@mail.usf.edu}}

\maketitle

\begin{abstract}
Offshoring the proprietary Intellectual property (IP) has recently increased the threat of malicious logic insertion in the form of Hardware Trojan (HT).  A potential and stealthy HT is  triggered with  nets that switch rarely during regular circuit operation. Detection of HT in the host design  requires exhaustive simulation to activate the HT during pre- and post-silicon. Although the nets with variable switching probability less than a threshold are primarily chosen as a good candidate for Trojan triggering, there is no systematic fine-grained approach for earlier detection of rare nets from word-level measures of input signals. In this paper, we propose a high-level technique to estimate the nets with the rare activity of arithmetic modules from word-level information. Specifically, for a given module, we use the knowledge of internal construction of the architecture to detect ``low activity'' and ``local regions'' without resorting to expensive RTL and other low-level simulations. The presented heuristic method abstracts away from the low-level details of design and describes the rare activity of bits (modules) in a word (architecture) as a function of signal statistics. The resulting quick estimates of nets in rare regions allows a designer to develop a compact test generation algorithm without the knowledge of the bit-level activity.  We determine the effect of different positions of the breakpoint in  the input signal to calculate the accuracy of the approach. We conduct a set of experiments on six adder architectures and four multiplier architectures. The average error to calculate the rare nets between RTL simulation and estimated values are below 2\% in all architectures.
\end{abstract}

 \section{Introduction} 
 
%{\textbf{HT (What is it, activation, motivation)}}: 

The use of fabrication equipment in offshore for manufacturing and testing Integrated Circuits (IC) has become common in the semiconductor design eco-system. In the long electronics supply chain with untrusted entities, IC has become prone to malicious modifications.  Various malicious manipulations (insertions or deletions) exist that modify part of the design  so that an attacker objective is achieved.  Such covert manipulations, known as Hardware Trojan (HT),  may affect the system by leaking the secret information, disabling parts of the system, weakening performance with early failures. HT is normally activated during a small time window and infrequent time of a circuit operation with low efforts.  Since HT is typically connected to the rare switching nets of  design, the time required to activate the triggering mechanism of HT can be significant. Various detection approaches exist that attempt to finding the minimal variations in power and timing due to the presence of HT. However, variants of HT concerning their physical properties, activation, and action characteristics make the current HT detection approaches non-unified \cite{5406669}.

%{\textbf{-- Detection technique (Current state / quick overview)-- drawbacks?}}: 

To be stealthy as possible, an attacker utilizes rare nets in  design to insert HT without any functional modification(s). Further, the size of the HT is adjusted (3-4x smaller than original design) accordingly so that any possible change in design parameter (timing, power, area) is insignificant during  post-silicon detection technique (e.g. side-channel analysis, SCA) \cite{4223234}. However, measurements during SCA are susceptible to pessimistic manufacturing variations. Several works studied how to improve the resolution of path-based timing analysis with additional test structure \cite{6394229, 6224324}, on-chip time-to-digital converter \cite{7827621}. On the contrary, logic-based detection requires switching activity analysis of the internal nets that facilitate a non-invasive technique to investigate  the possible HT in a design. This analysis acts as the de-facto for improved test vector generation to detect HT during pre-silicon \cite{Banga:2008:GTG:1366110.1366196, dupuis2013identification} and post-silicon \cite{7059101, salmani2012layout} using functional testing and verification.

Moreover, the switching activity of the design (both HT-free and HT-affected) is input vector dependent which can be generated randomly or following spatial correlation \cite{8607170}.  Further, functional testing is independent of process parameters that make it suitable for the attacker to simulate and find out the rare switching nets during pre-silicon. As the switching activity of the nets can vary within a wide range of values, an intelligent attacker can find a way of combining the rare nets with variable toggle rate to trigger HT. In both cases, the analysis of switching activity file regarding input vector occurs before HT insertion or detection.  Such activity analysis  also avoids the expensive and destructive de-packaging and de-layering of the encapsulated IC.

In this paper, we propose a new approach for estimating nets  that switch rarely from input signal word-level statistics in a given Register Transfer Level (RTL) description. High-level estimation of rare nets provides fast and efficient localization of internal signals within an arithmetic module that can be responsible for hard-to-detect HT activation. Given a technology-independent RTL description of the arithmetic module, the proposed technique will guide the designer (defender)  to locate  rare transition activity nets analytically  which is dependent on word-length and its' statistics. This information can be used for efficient segmentation of a module into smaller sub-module(s)  before logic synthesis. We develop this characterization technique based on Dual-Bit-Type (DBT) model \cite{386219} of the datapath components (adder and multiplier). The model breaks the component word-length into three regions of continuous bits: LSB, linear, and MSB regions. Highly correlated bits  are found in MSB regions that exhibit non-random behavior and low transition activity. Thus, transition activity  at MSB node(s) of the primitives (adder, multiplier, register, etc.), has been used to provide rare activity estimates of the architecture.

An attacker normally chooses nets with very rare internal logic conditions (low controllability and observability) to trigger  HT. For efficient activation of HT, attacker considers the region(s) with low bit-level activity to avoid accidental triggering. Analytically, with known delay and statistical distribution of the architecture, an IP integrator can distinguish between competing architectures in terms of the lower rarity nets from word-level  characteristics.  To the best of our knowledge, the proposed approach is first to identify modules whose models of rare activity are characterized by statistics of input word-lengths. In summary, the novelty and contributions of the paper are as follows:
\begin{itemize}
    \item high level modeling of rare activity nets and location of these nets in the arithmetic module.
    \item application of word-level statistics (mean, variance, and correlation coefficient) to estimate rare nets and hence complementing expensive simulation.
    \item technology independent, closed-form analytical techniques to estimate  rare nets in MSB region(s).
\end{itemize}

The rest of the paper is organized as follows. Section \ref{sec:back} provides background on HT detection techniques based on switching activity analysis. Section \ref{sec:proposed} describes the attack model, theoretical background and framework to estimate rare nets. Section \ref{sec:results} reports the experimental results. Finally, section \ref{sec:conclude} draws the conclusion and future work.

\section{Background and Related Work}
\label{sec:back}

We briefly summarize the methods to detect (and possibly remove) malicious functionality at behavioral, RT-, and gate-level design. In particular, we review only the compact test vector generation technique to identify rare nodes during functional testing under specific assumptions and search space. A comprehensive overview of HT diversity is available in \cite{Xiao:2016:HTL:2948199.2906147}.

Although the HT triggering mechanisms are non-trivial, we  classify the test pattern generation methods into two broad categories, namely, statistical- and  probabilistic-modeling. With statistical modeling, one can simulate the circuit under randomly generated test patterns and differentiate rare nets from the rest based on an arbitrary triggering threshold. Statistical technique such as MERO \cite{10.1007/978-3-642-04138-9_28} prunes the test vector space functional simulation to improve `Trojan Coverage' and `Trigger Coverage'. Genetic algorithm and Boolean Satisfiability based `Trojan sensitization' have been proposed in  \cite{cryptoeprint:2015:1252} to improve the detection sensitivity. An automatic and compact test vector generation algorithm to aid in the side-channel analysis is proposed in \cite{8342270, Huang:2016:MST:2976749.2978396}. An improved methodology to increase bit-level transition activity is proposed in \cite{7082769}. These approaches did not take data correlation (both spatial and temporal) into account and utilized  random vector-based simulation. However, the probabilistic
model of dependencies exists for input data sequence that may appear at the re-convergent input(s) in a design \cite{1600469, 681258}. 

Probabilistic modeling approaches propagate the switching probability of primary inputs to estimate the internal switching activities in  design. Characteristic polynomial based signature of the Circuit Under Test (CUT) is proposed in \cite{4708870}. With the help of 2-to-1 MUXs, the switching activity of rare nets has been improved considering only one form of transition ($1 \rightarrow 0$ or $0 \rightarrow 1$) \cite{6971844}.

Statistical signal correlation-based HT detection techniques avoid the triggering state and payload sensitization at the output. In \cite{Cakir:2015:HTD:2755753.2755860}, the authors presented an information-theoretic approach  to simulation data to detect HT. Cross-correlation based test vector generation technique for the hard-to-reach region in  design is proposed in \cite{8567415}.

Unlike previous studies, our work (a) focuses on word-level statistical behavior to estimate rare nets, (b) provides an early-on estimation framework without RTL and low-level simulation, and (c) requires no knowledge of internal implementation of the architecture.

\section{Proposed Approach and Implementation}
\label{sec:proposed}

% \textbf{Theoretical high-level modeling for spatio-temporal corr?} \\
% \textbf{How we propagate statistical param. for RTL components?} \\
% \textbf{How we find rare nets for RTL components?}\\

\subsection{Threat Model}
\label{threat_model}

\begin{figure}
\centering
\includegraphics[width=\columnwidth]{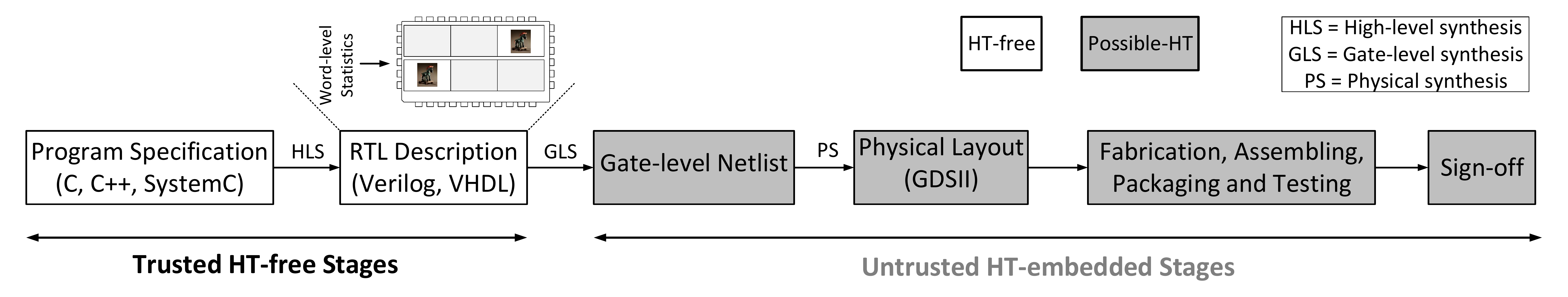}
\caption{Attack model for HT vulnerability analysis in RTL designs. Trojan symbol is reproduced from \cite{mitra2015stopping}.}
\label{fig:attack_model}
\vspace{-4ex}
\end{figure}

Traditional HT attacks deal with an agent who maliciously inserts some of her chosen logic  to ensure that the additional circuitry will be activated during rare conditions. In our threat model, we assume two parties, a benign designer, who develops the RTL model using trustworthy High-Level Synthesis (HLS) tool and an attacker to whom the designer ports the RTL design for system integration during pre-silicon or from whom the end-user accepts the design as a packaged product during post-silicon (Fig. \ref{fig:attack_model}). We also assume the attacker has access to a subset of IP models drawn from the same distributor to make it harder to discriminate between malicious and benign IP. As part of the attacker objective, they control exactly the triggering logic and location of HT that have minimal impact on global parameters (power, performance, and area). Another goal of the attacker is to ensure the higher misclassification rate against the measures to detect HT. We can broadly classify attacker goal in two categories (targeted and non-targeted). In a targeted attack setting, h/she may disable the device on-field or degrade the reliability earlier than Mean Time to failure (MTF). During a non-targeted attack, the adversary may aim to leak sensitive information as a backdoor instead of `visible' consequences. 

We aim  to explore the possibility and location of HT attack in an RTL IP using its model parameters. Given a high-level IP/IC description, our approach checks two critical properties. First, the designer can locate whether a module in IP, in isolation, generates more rare activity nets than others given a triggering threshold. Second, a compact test pattern generation algorithm can be developed to identify any malicious updates. We also find that the proposed analytical approach can be complemented with an expensive RTL simulation to provide a qualitative notion of stealthy HT behavior.

\subsection{Theoretical modeling for transition activity estimation}
\label{input_cor}

Let $X_{t}^{N}$ be an $N$-bit signal in the time interval $(-\frac{T}{2}, \frac{T}{2}]$ for a single-input module. Given normal distribution of input data environment, the signal probability for the $i$-th bit, $N_{i}$ of  $X_{t}^{N}$ to be evaluated to logic-1 can be calculated as follows \cite{644033}: 
\begin{equation}
p_{i} = Pr (N_{i} = 1) = \sum_{\forall x \in \chi_{i}} \frac{1}{{\sigma \sqrt {2\pi } }}e^{{{ - \left( {x - \mu } \right)^2 } \mathord{\left/ {\vphantom {{ - \left( {x - \mu } \right)^2 } {2\sigma ^2 }}} \right. \kern-\nulldelimiterspace} {2\sigma ^2 }}}
\end{equation}
%\cite{386219} 
where $\chi_{i}$ is the set of all elements in $\chi$ that the signal $X_{t}^{N}$ can assume. The value $p_{i}$ at any net can be derived, given word-level statistical parameters such as mean ($\mu_{X}$), variance ($\sigma^{2}_{X}$), and spatio-temporal autocorrelation ($\rho_{X}$). Therefore, temporally uncorrelated input data leads to an error in the signal activity estimation of internal nets. Mean, variance, and spatio-temporal autocorrelation of $X_{t}^{N}$ can be expressed as 
\begin{equation}
\label{eq:mean}
\mu_{X} = E [X_{t}^{N}].
\end{equation}

\begin{equation}
\label{eq:sigma}
\sigma_{X} = \sqrt{E[X_{t}^{2}] - E^{2}[X_{t}]} = \sqrt{E[X_{t}^{2}] - \mu^{2}} = p_{i} - p_{i}^{2}
\end{equation}

\begin{equation}
\label{eq:rho}
\rho = \frac{E(X_{t}^{N},X_{t-1}^{N}) - \mu_{X}^{2}}{\sigma_{X}^{2}} = \frac{cov(X_{t-1}^{N}, X_{t}^{N})}{var(X^{N})}  
\end{equation}

% %\cite{892863}
The normalized transition activity (toggle), $\alpha$ of the $X_{t}^{N}$ over all bit positions is given by 
\begin{multline} 
\label{eq:normalized tc}
\alpha = \sum_{i=0}^{N-1}P(\{X^{N}(t-T)\bar{X}^{N}(t)\} \cup \{\bar{X}^{N}(t-T)X^{N}(t)\}) \\
        =  2\sum_{i=0}^{N-1}p_{i}(1-p_{i}) 
        = \sum_{i=0}^{N-1} N_{i}
\end{multline}

% %\cite{644033}
where $X^{N}(t-T)\bar{X}^{N}(t)$ denotes a logic-1 to logic-0 transition, $\bar{X}^{N}(t-T)X^{N}(t)$ denotes a logic-0 to logic-1 transition, and T is the clock period. We can also define $\alpha$ from \cite{705205} to input data for exact synthesis of single-bit signal in terms of $\rho$ as follows:
\begin{equation} 
\label{eq:singlebit tc}
\alpha = 2\sum_{i=0}^{N-1} p_{i} (1-p_{i}) (1-\rho_{i})
%\vspace{-4ex}
\end{equation}

% %DBT Model

\begin{figure}
\centering
\includegraphics[width=\columnwidth]{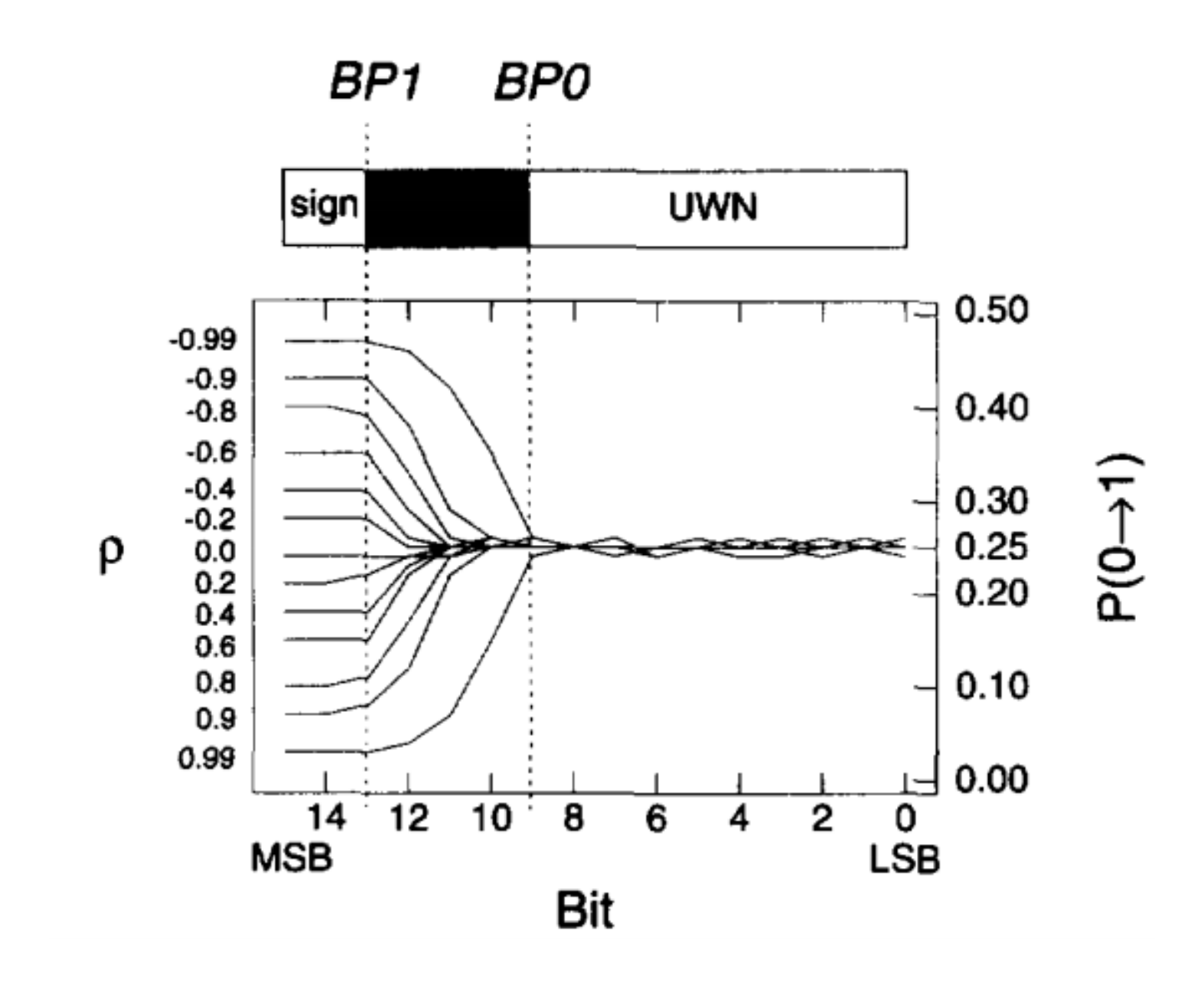}
\vspace{-3ex}
\caption{Transition activity of a signal bits for white Gaussian process with 16-bit two's complement form and varying temporal correlation (reproduced from \cite{386219}). }
\label{fig:Landmanfig3}
\vspace{-4ex}
\end{figure}

For uncorrelated data $(\rho_{i} = 0)$, Eqn. \ref{eq:singlebit tc} reduces to Eqn. \ref{eq:normalized tc}. For a given zero-mean Gaussian signal, the signal probability ($p_{i}$) at the $i^{th}$ bit position is 0.5 in 2's complement representation. Hence, bit-level switching activity in Eqn. \ref{eq:singlebit tc} can be rewritten as follows:
\begin{equation} 
\label{eq:indiv_tc}
\alpha_{i} = 0.5(1-\rho_{i})
%\vspace{-4ex}
\end{equation}

An accurate estimation of switching activity (independent of encoding of the signal) in the sign-bit, $\alpha_{msb}$  has been proposed in \cite{705205} in terms of Eqn. \ref{eq:rho}:
\begin{equation} 
\label{eq:alpha_msb}
\alpha_{msb} = \frac{1}{\pi} cos^{-1}(\rho)
%\vspace{-4ex}
\end{equation}

From Eqns. \ref{eq:indiv_tc} and \ref{eq:alpha_msb}, we can determine the correlation of sign-bit $\rho_{msb}$ in terms of word-level correlation ($\rho$) from the following expression:
\begin{equation} 
\label{eq:row_msb}
\rho_{msb} = \frac{2}{\pi} sin^{-1}(\rho)
%\vspace{-4ex}
\end{equation}

{\textbf{Calculation of BP$_{0}$ and BP$_{1}$}}: Using a computationally inexpensive method, we can divide a signal in 2's complement representation into three different regions (LSB, linear, and MSB) based on transition activity.  We can see from  Fig. \ref{fig:Landmanfig3} that the temporal correlation ($\rho$) from the LSB up to a first breakpoint $BP0$ is almost zero and hence we observe the maximum switching activity in LSB region ($0 \leq i \leq BP_{0}$). The uncorrelated bits in LSB region exhibit random switching where both $p_{i}$ and $\alpha_{i}$ are equal to $\frac{1}{2}$.  We can see a linear increase in $\rho$ from $BP0$ to sign bit (MSB) and lower switching activity in the sign region. In the linear region ($BP_{0} \leq i \leq BP_{1}$), we see an increasing spatial correlation and correspondingly, decrease in the switching activity.
 We can compute the first breakpoint, BP$_{0}$ as follows from \cite{757374}:
\begin{equation} 
\label{eq:bp0}
BP_{0} = \nint{log_{2}[2\sigma(1 - \rho_{msb})]}
%\vspace{-4ex}
\end{equation}
where $\nint{x}$ is the rounding operation. We multiply by two to include both positive and negative region (-2$^{N-1}$ $\leq$ $X_{t}^{N}$ $\leq$ 2$^{N-1}$ - 1). Similarly, we can define BP$_{1}$ as follows from \cite{757374}:

\begin{equation} 
\label{eq:bp1}
\begin{split}
BP_{1} &= \nint{log_{2}[({X_t}_{max}^{N} - {X_t}_{min}^{N})\sqrt{(1 - \rho_{msb})}]} \\
&= \nint{log_{2}[(\mu_{x} + 3\sigma_{x} - \mu_{x} + 3\sigma_{x})\sqrt{(1 - \rho_{msb})}]} \\
&= \nint{log_{2}[6\sigma_{x}\sqrt{(1 - \rho_{msb})}]}
\end{split}
\end{equation}

Hence, with the knowledge of BP$_{0}$ and BP$_{1}$, we can express the correlation coefficient values as follows \cite{644033}:
\begin{equation}
\label{row_all}
  \rho_{i}=\begin{cases}
    0, & \text{($i < BP_{0}$)} \\
    \frac{\rho_{BP_1}(i - BP_{0}+1)}{BP_{1} - BP_{0}}, & \text{($BP_{0} \leq i \leq BP_{1} - 1 $)} \\
    \rho_{msb}, & \text{($i \geq BP_{1} - 1 $)} 
  \end{cases}
\end{equation}

We can also approximate the switching activity model of $i^{th}$ bit using Eqn. \ref{row_all} as follows \cite{822629}:
\begin{equation}
\label{switch_all}
  \alpha_{i}=\begin{cases}
    2p_{i}^{1}(1 - p_{i}^{1}), & \text{($i \leq BP_{0}$)} \\
    0.5 + (\alpha_{msb} - 0.5)\frac{i - BP_{0}}{BP_{1} - BP_{0}}, & \text{($BP_{0} < i < BP_{1}$)} \\
    \alpha_{msb}, & \text{($i \geq BP_{1}$)} 
  \end{cases}
\end{equation}

\subsection{Framework for rare activity nets estimation}
\label{rare_net_est}

\begin{figure}
\centering
\includegraphics[width=\columnwidth]{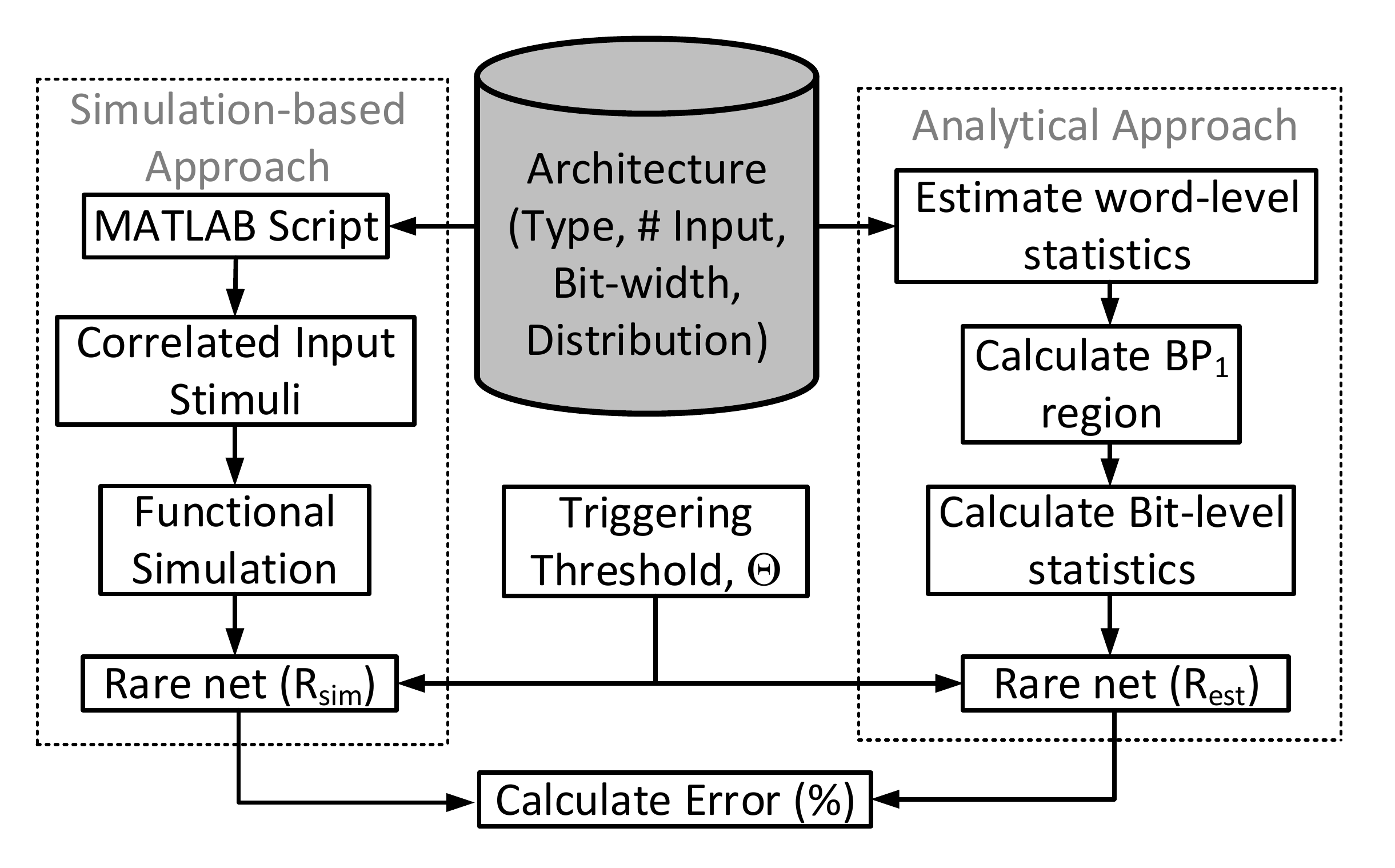}
\vspace{-3ex}
\caption{Framework for rare nets estimation and  error calculation between simulation and analytical approach.}
\label{fig:framework}
\vspace{-4ex}
\end{figure}

As we can see from Fig. \ref{fig:Landmanfig3}, the highly correlated bit(s) in the MSB region of a word lead to minimal switching. As such, the MSB region(s) manifest themselves as a possible location of HT. Alternatively, 
if higher toggle activity of some internal nets happens in rare condition, one can focus on LSB region of a signal word. The length of both regions can be demonstrated from word-level models of macro-blocks. The high-level estimation flow to identifying and localizing Trojan vulnerable blocks using model parameters is shown in  Fig. \ref{fig:framework}. It  contains three steps, namely, modeling, estimation, and simulation phase.

\textbf{Modeling phase}: We assume the signal shows Gaussian distribution  with non-zero mean and  is in two's complement  representation. The modeling parameters selected are independent of the distribution type and signal representation.   On a higher level of abstraction, we can model the word-level characteristics as follows:
\begin{equation} 
\label{eq:model}
(\mu, \sigma, \rho)_{A} = f (X_{t-1}^{N}, X_{t}^{N}, BW) 
\end{equation}

where $X_{t-1}^{N}$ and $X_{t}^{N}$ refer to the signal into consecutive timestamp and $BW$ being the input bit-width of the signal. We can determine $\mu_{A}$ and $\sigma_{A}$ from $BW$ of the signal using Eqns. \ref{eq:mean} and \ref{eq:sigma}.  Using these characteristics data, we can determine the bit-level statistics of any RTL design. The bit-level characteristics can be used to derive an exact estimation of rare nets that are captured during input dependent functional simulation of the design. In this paper, we restrict our analyses to word-level information  and it provides us architecture- and pattern-independent tight upper bound estimation of the nets that fall under particular triggering probability. In some cases, the accuracy loss can be significant which is architecture-dependent but it can be significantly improved by considering bit-level statistics at the cost of characterization time.  Nevertheless, our word-level modeling approach can be used for architecture's characterization of HT vulnerability in less than a minute. We present a heuristic to intelligently select sub-module(s) that have shown good accuracy in practice.

\textbf{Estimation phase}: Enumerating all possible input patterns (4$^n$ patterns for $n$ primary inputs) for a large circuit are not helpful to guide us the Trojan location and triggering logic.
During the simulation-based approach, an attacker  considers the sign transition(s) at the internal nodes of a module which are input pattern dependent and drawn from a particular distribution. It also turns out that the inexact delay model during simulation can lead to inaccurate transition probability at circuit nodes.

On the contrary, the statistical properties of the possible input stream in a design can lead to better search technique of Trojan location. Moreover, the statistical estimation can ignore the delay influences. Although the assumption made by the attacker on the signal distribution cannot be assured during pre-silicon, word-level statistical information required for breakpoint estimation is independent of the distribution type. To reduce the complexity and higher flexibility in the estimation, we focus on calculating breakpoints (BP$_{0}$ and BP$_{1}$) from statistical properties. Similar to Eqn. \ref{eq:model}, we can determine a functional relationship between  breakpoints, signal statistics, and architecture bit-width as follows:

\begin{equation} 
\label{eq:est_bp}
(BP_{0}, BP_{1})_{arc} = f (X_{t-1}^{N}, X_{t}^{N}, BW) 
\end{equation}

where ($X_{t-1}^{N}, X_{t}^{N}$) and $BW$ will provide  the required $\rho_{msb}$ to determine breakpoints from Eqns. \ref{eq:bp0} and \ref{eq:bp1}. Given an RTL datapath design, we divide the circuit into a set of sub-components where each sub-component is a bit-slice design.  Considering each sub-component separately, we can find total nets  from the structural description of the architecture. Using Eqn. \ref{eq:est_bp} for breakpoints estimation, first, we find the sub-module(s)  that lie from BP$_{1}$ position to the largest bit position required to represent the signal and then the nets within  these sub-modules to calculate total rare triggering nets. Let us assume there are $m$ modules in the architecture of type $i$, $1 \leq i \leq m$ and each type has $n$ nets. If the set of modules that can be responsible for providing rare nets to Trojan triggering are $j$ ($1 \leq j \leq i$), the following equation accounts all these rare nets:
\begin{equation} 
\label{eq:est_nets}
T_{rare} = R(m,i,j) = \sum_{j=1}^{} j*m_{j}
\end{equation}

Clearly, the module having least rare nets can be found as:
\begin{equation} 
\label{eq:least_nets}
T^{least}_{rare} = \min_{i \in m} R(m,i,j)
\end{equation}

\textbf{Simulation phase}: To investigate the model accuracy,  we perform gate-level simulation and measure the difference of nets between estimated and simulated value. For the simulation, we generate the correlated input stream according to statistics of the above modeling and estimation phase. For each n-bit arithmetic architecture, we generate different word-level statistics ($\mu, \sigma, \rho$) and calculate the breakpoints (BP$_{0}$, BP$_{1}$). Depending on the statistics, the signal value can range from ($\mu_{x} - 3\sigma_{x}$) to 
($\mu_{x} + 3\sigma_{x}$). For each choice of the breakpoint (BP$_{1}$), we perform the simulation to count the nets whose signal transitions fall under particular triggering probability. For  architecture  with two operands and (un)equal bit-width, we can determine the upper  and lower bound in the LSB and MSB region as follows \cite{4541787}:

\begin{equation} 
\label{eq:min_max_bp}
\begin{split}
BP^{min}_{0} &= min (A_{BP_{0}}, B_{BP_{0}}) \\
BP^{max}_{1} &= max (A_{BP_{1}}, B_{BP_{1}})
\end{split}
\end{equation}

To assess the model, we use the following equations to estimate absolute error, $e$ and mean square error, $\bar{e}$ :
\begin{equation} 
\label{eq:error}
\begin{split}
e &= \abs{\frac{P_{sim} - P_{est}}{P_{sim}}} \\
\bar{e} &= \frac{1}{n}\sum_{i=1}^{n} e_{i}
\end{split}
\end{equation}

where $n$ denotes the number of triggering threshold bound in a particular BP$_{1}$ position, $P_{sim}$ and $P_{est}$ refer the simulated and estimated rare triggering nets.

\begin{figure}
\centering
\includegraphics[width=\columnwidth]{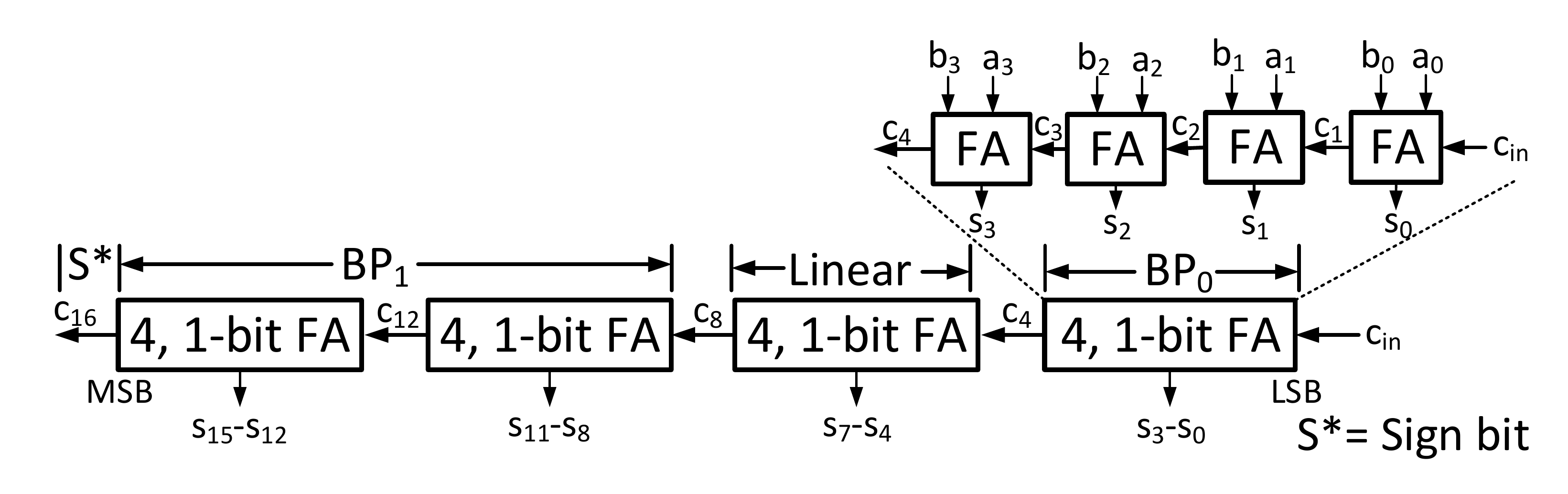}
\vspace{-3ex}
\caption{Decomposition of 16-bit Ripple-Carry Adder into three regions according to DBT \cite{386219}.}
\label{fig:example_mot}
\vspace{-5ex}
\end{figure}

\begin{figure}
\centering
\includegraphics[width=\columnwidth]{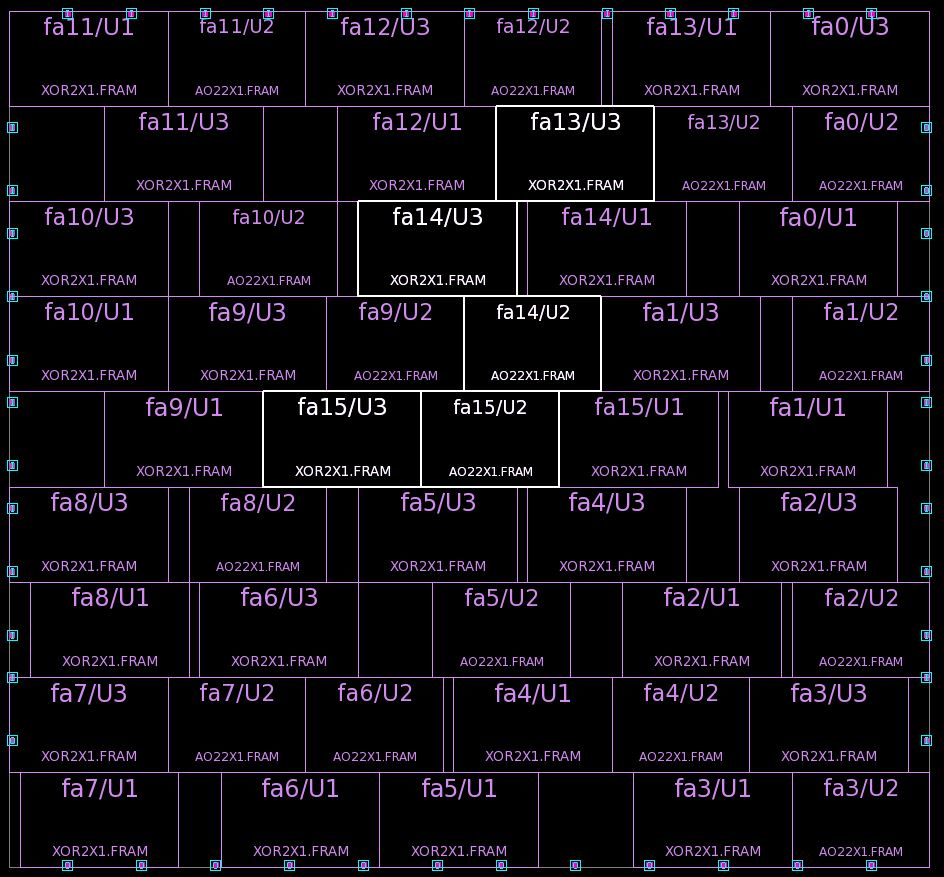}
\vspace{-3ex}
\caption{Floor-planning and placement of  16-bit RCA with leaf cells highlighted as the location of rare triggering nets.}
\label{fig:pnr_rca}
\vspace{-4ex}
\end{figure}

\textbf{Motivational example to estimate rare activity nets}: A 16-bit Ripple-Carry Adder (RCA) is presented in Fig. \ref{fig:example_mot} where we decompose the adder into four blocks. Each block contains four, 1-bit Full Adder (FA). Let us assume, given the statistics in terms of input operands, we calculate the breakpoints position as  BP$_{0}$ (0$^{th}$ to 3$^{rd}$), linear (4$^{th}$ to 7$^{th}$), and BP$_{1}$ (8$^{th}$ to 15$^{th}$). As mentioned earlier, the rarest activity will be generated in the BP$_{1}$ region. Hence, HT vulnerable region can be modeled as the location of FA from 15$^{th}$- to 8$^{th}$-bit position and the sum of nets in these FA's in 3$^{rd}$ and 4$^{th}$ block constitute the upper bound of rare activity nets for ripple-carry adder. However, in the case, where the operands have different bit-width, we see two distinct scenarios. In one case, B$_{BP1}$ is contained in A$_{BP0}$  when (A$_{BP0}$ > B$_{BP0}$) and vice-versa when (A$_{BP0}$ < B$_{BP0}$). In both cases, we use Eqn. \ref{eq:min_max_bp} to calculate the breakpoints.  For triggering threshold < 10$^{-5}$, we found 5 nets that belong to FA$_{13}$, FA$_{14}$ and FA$_{15}$. After the module placement shown in Fig. \ref{fig:pnr_rca}, we also found the geometric positions of these cells are significantly close enough  to localize HT triggering signal that would increase HT impact. Hence, the word-level estimation clearly indicates that the majority of rare transitions happen in BP$_{1}$ position.

\section{Experimental Results}
\label{sec:results}

% \textbf{Generic data distribution based simulation results vs. modeling result for different RTL components?}

In this section, we present the results of our word-level statistics based on rare activity net modeling approach. We have evaluated the accuracy of the approach on six adder and four multiplier architectures each having width of 8- and 16-bit. For each architecture, we assume two operands are available with equal bit-width. All architectures are taken from OpenCores (cite).

First, we generate correlated input vectors for different BP$_{1}$ positions using an in-house MATLAB script. Then we perform RTL simulation using Synopsys VCS-MX on each architecture for 10000 input vectors and find out the nets having variable toggle probability in between 0 to 10$^{-6}$. The total number of rare nets within a toggle threshold by the analytical approach of a given architecture are compared to those from the simulations and accordingly, the average errors are computed.

%%%%%%%%%%%%%%%%%%%%%%%%%%%%%%%%%%%%%%%%%%%%%%%%%%%%%%%%%%%%%%%%%%
% Merged figures Adder
\begin{figure*}[tb]
\centering
\resizebox{\textwidth}{!}{
\begin{tabular}{c c c c}
\includegraphics[width=1.1\textwidth]{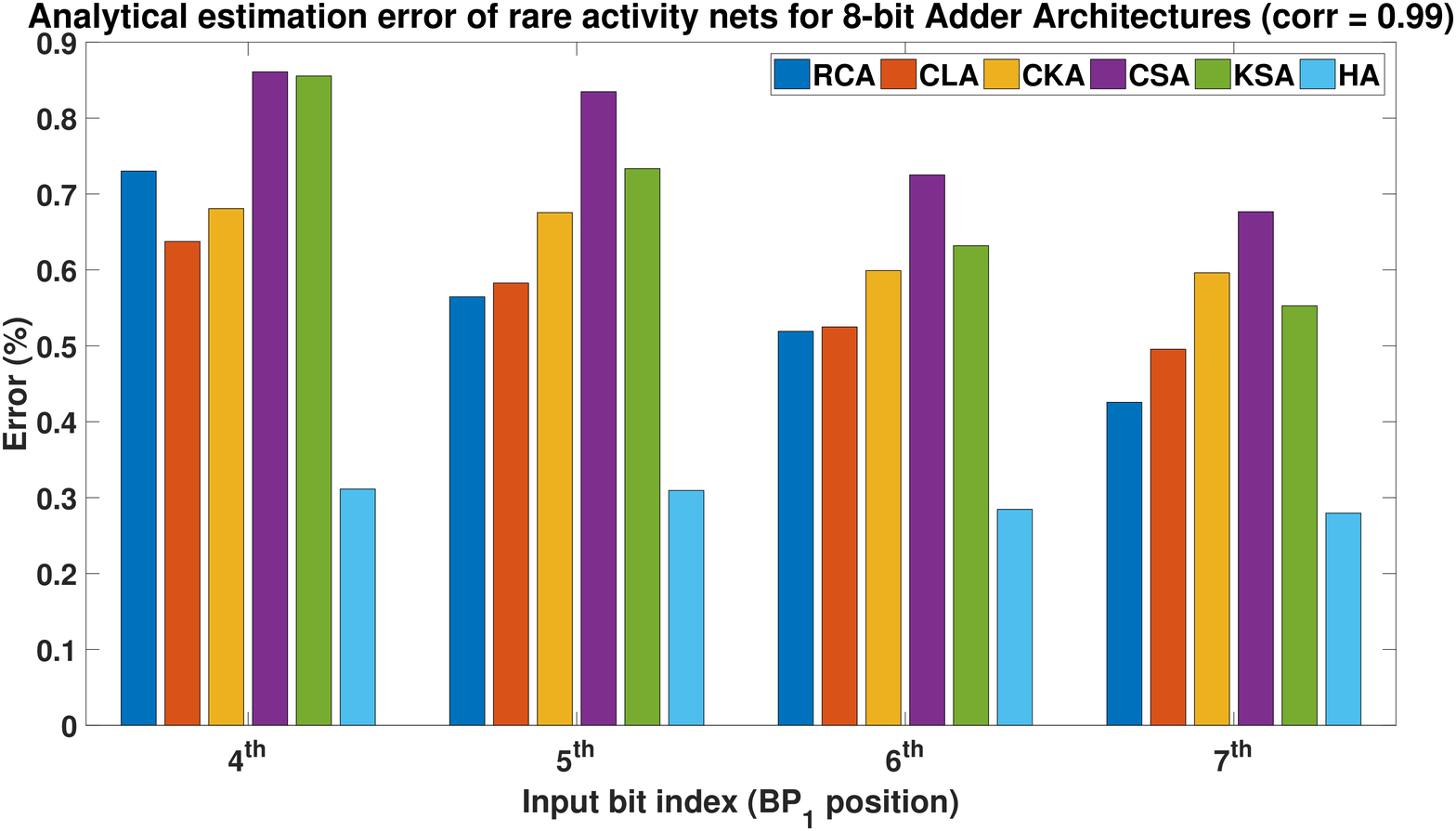} & 
\includegraphics[width=1.1\textwidth]{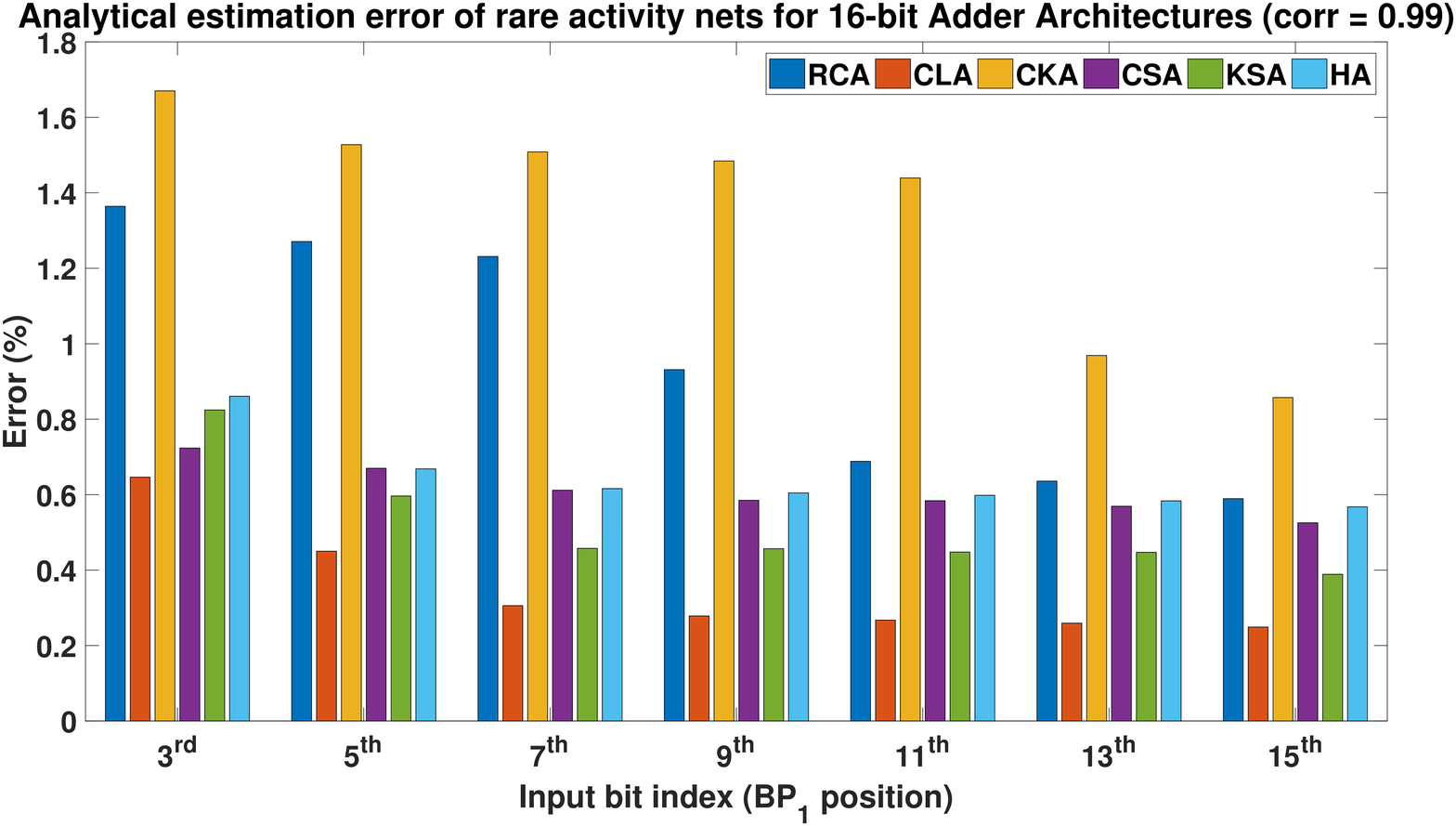} & 
\includegraphics[width=1.1\textwidth]{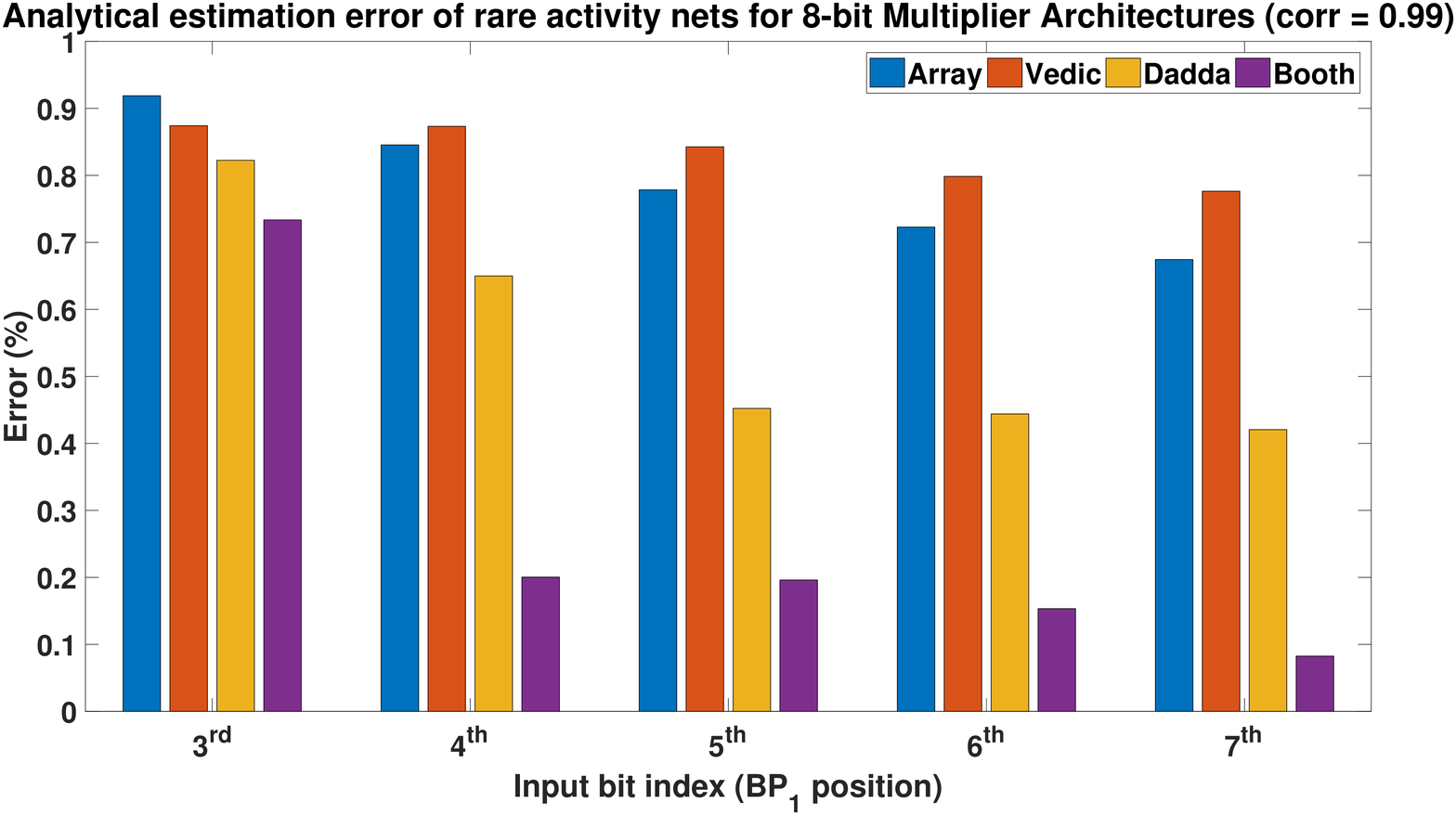} & 
\includegraphics[width=1.1\textwidth]{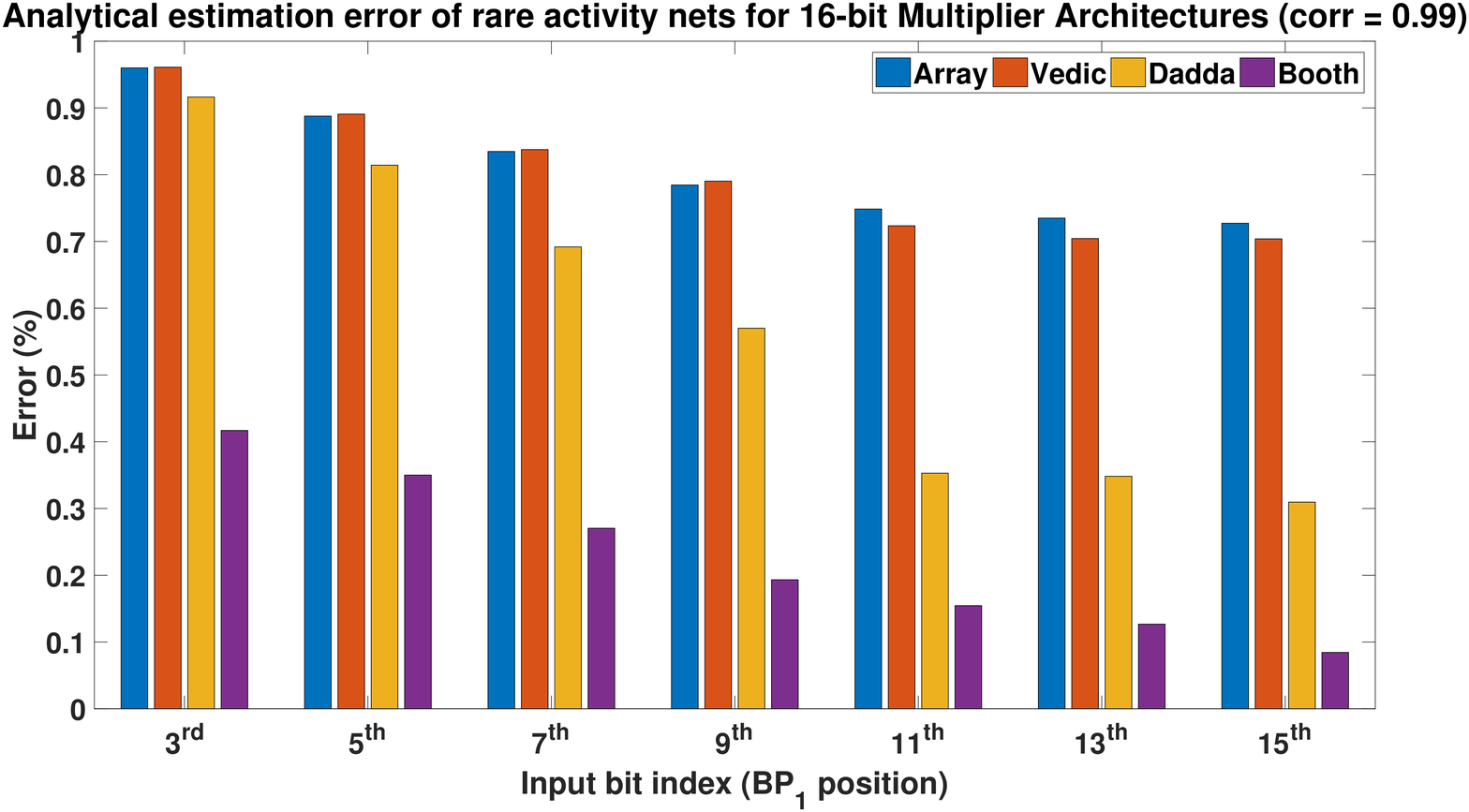} \\
\textbf{\LARGE (a)} & \textbf{\LARGE(b)}  & \textbf{\LARGE(c)} & \textbf{\LARGE(d)} \\
\end{tabular}}
\caption{Comparison of accuracy estimation in terms of rare nets for adder and multiplier architectures.}
\label{fig:adder_plot}
\vspace*{-4ex}
\end{figure*}

\textbf{Analytical estimation error of adder architectures:} We have considered Ripple Carry (RCA), Carry Lookahead (CLA), Carry Skip (CKA), Carry Select (CSA),
Kogge-Stone (KSA), and Hybrid Adder (HA). Fig. \ref{fig:adder_plot} (a,b) shows error (\%) vs. BP$_{1}$ positions for adders of 8- and 16-bit for correlation ($\rho$ = 0.99). Reference values of rare nets are obtained from Switching Activity Interchange Format (SAIF) file after the RTL simulation. It is evident from the figure that as BP$_{1}$ moves towards sign bit region, analytical estimation is close to functional simulation. One important source of error is when BP$_{1}$ is close to LSB region. This is because we see a limited range of random activity and simulation also considers the glitch activities (e.g. $0 \rightarrow X \rightarrow 1$ or $1 \rightarrow X \rightarrow 0$) as a transition. An attacker may want to localize the triggering signals within a sub-module to evade SCA. Otherwise, incorporating  rare nets from different sub-modules that are placed far away in architecture would increase detection sensitivity.

From Table \ref{tab:avg_err}, we  see CLA shows average error less than 0.4\% while that due to CKA, the average error is around 1.35\%. This is because, in CLA, we have unrolled carry equations to build carry network of given bit-width. Although the delay increases due to each additional level of lookahead, however, we can closely approximate the rare nets as we decompose the carry equations to basic gates.  In CKA, we have skip paths for each 4-bit adder blocks. Although we have the shortest carry propagation time through the skip blocks, we even include the nets of the skip logic when the bit-width from BP$_{1}$ position to sign-bit is not multiple of 4. Though it simplifies the estimation, it implies an effect on error calculation. 

\begin{table}
\begin{center}
\caption{Average estimation error (\%) for different BP$_{1}$ positions in a given architecture and bit-width.}
\label{tab:avg_err}
{\begin{tabular}{|c|c|c|c|c|c|c|}
\midrule
 & Arch. & 8-bit & 16-bit  \\ \midrule
\multirow{6}{*} {Adder} & RCA & 0.95  & 0.55\\
& CLA & 0.56 & 0.35\\
& CKA &  1.35 & 0.63\\ 
& CSA &  0.77 & 0.60\\ 
& KSA &  0.69 & 0.51\\ 
& HA & 0.64 &0.29 \\ \cline{1-4}
\multirow{4}{*} {Multiplier} & Array & 0.81 & 0.78\\
& Vedic & 0.83 & 0.80 \\
& Dadda &  0.57 & 0.55\\ 
& Booth & 0.27 & 0.22\\  \cline{1-4}
\end{tabular}}
\end{center}
\vspace{-6ex}
\end{table}

%\fbox{show result for other corr}

\textbf{Analytical estimation error on multiplier architectures:} Similar to adder architecture, we consider four multiplier architectures (array, vedic, dadda, booth) of two different bit-widths (8- and 16-bit). We consider the correlation value ($\rho$) of the signal to be 0.99 and correspondingly we estimate the BP$_{1}$ position. For both bit-widths, we see booth multiplier shows least error (0.27\% for 8bit and 0.22\% for 16bit) whereas vedic multiplier shows the highest error (0.83\% for 8bit and 0.80\% for 16bit). For booth multiplier, the implementation is fully parallel and carry-free, hence the estimation closely matches with the simulation. For vedic multiplier, we can determine the partial products in parallel but it requires more than two additional levels of adders (e.g. CLA). These additional levels would sufficiently relate to the error (< 1\%) in vedic multiplier.

\section{Conclusion}
\label{sec:conclude}

In this paper, we present macro-models to estimate rare nets in adder and multiplier architectures using word-level input statistics. We have shown that input statistics can closely approximate the rare triggering probabilities of internal nets in design and locate them as well. Such modeling techniques of high-level rare activity nets can reduce the Trojan detection time and  complement the expensive low-level simulations. We analyzed both architectures of different bit-widths and found the error within 1-2\%. In  the future, we plan to find both combinational and sequential Trojan triggering logic from the  modeling-based approach with low false positive/negative rates.

%

%\nocite{*}
%\bibliographystyle{IEEE}
\bibliographystyle{unsrt}
\scriptsize{
\bibliography{bib/final.bib}
}

% \newpage
% \input{6_miscel.tex}

\end{document}